# Finite-Size Effect of Hadron Masses with Kogut-Susskind Quarks *

S. Aoki[a], T. Umemura[a], M. Fukugita[b], N. Ishizuka[a], H. Mino[c], M Okawa[d] and A. Ukawa[a]

[a]Institute of Physics, University of Tsukuba, Tsukuba, Ibaraki 305, Japan

[b]Yukawa Institute for Theoretical Physics, Kyoto University, Kyoto 606, Japan

[c]Faculty of Engineering, Yamanashi University, Kofu 400, Japan

[d]National Laboratory for High Energy Physics(KEK), Tsukuba, Ibaraki 305, Japan

We present numerical results and their analyses of finite-size effects of hadron masses for both quenched and full QCD calculations. We show that they are much larger for full QCD due to dynamical sea quarks and the associated breaking of $Z(3)$ symmetry. We also argue that finite-size effects are non-negligible even for the largest lattice size simulation currently being made for a very small quark mass.

## 1. Introduction

In this report we summarize a comparative study of finite size effects for hadron masses in quenched and full QCD with the Kogut-Susskind quark action[1]. Our motivation for the study is, first, to examine whether the impression obtained from previous quenched studies that finite-size effects are perhaps smaller for quenched QCD[2] is actually valid, and, second, to understand the origin of the difference if it exists, including the possible role of dynamical sea quarks in the large finite-size effects observed for full QCD[3].

Our two-flavor full QCD study[3] was carried out at $\beta = 5.7$ for the spatial lattice size in the range $La = 0.7 - 1.8$fm ($L = 8 - 20$) with $a = 0.089(3)$fm fixed by the $\rho$ meson mass. For a comparative quenched study we chose $\beta = 6.0$ where our previous calculation on a $24^3 \times 40$ lattice gave $a = 0.105(3)$fm[4], and made new runs for the sizes $L = 6 - 16 (La = 0.63 - 1.7$fm$)$. We also added a full QCD run on an $8^3 \times 16$ lattice using quark boundary conditions different from those of Ref. [3].

## 2. Comparison of lattice-size dependence

Our full and quenched QCD results for $\pi$, $\rho$ and $N$ masses are compared in Fig. 1 where the periodic boundary condition is imposed on sea and valence quarks in the spatial directions. It is evident in Fig. 1 that the magnitude of finite-size effects is much smaller for quenched QCD than for full QCD, especially below $La \simeq 1$ fm. The size dependence is also significantly weaker for quenched QCD; assuming a power law $\delta_L m \propto L^{-\alpha}$ we find $\alpha \approx 1 - 2$ for quenched QCD as compared to $\alpha \approx 2 - 3$ observed for full QCD[5].

## 3. Origin of difference

The difference observed in Fig. 1 can be understood by a simple argument based on an expansion in inverse powers of quark masses. The essential points of the argument are present in the literature, dating back to quite early times for the quenched case[7] and more recently for full QCD[5].

For meson propagators application of the expansion for valence quarks leads to an expression of form

$$\sum_C m_{val}^{-l(C)} \cdot \langle W(C) \rangle + \sum_C m_{val}^{-l(C)} \cdot \sigma_{val} \cdot \langle P(C) \rangle, \quad (1)$$

where $m_{val}$ is the valence quark mass and $\langle \cdot \rangle$ denotes the gluon field average (including the quark determinant in full QCD). The second term comes from valence quark loops $C$ wrapping around the lattice in the spatial directions (Polyakov-type), and the first from trivial ones (Wilson-type), with $P(C)$ and $W(C)$ the associated gauge link factors. The sign factor $\sigma_{val}$ for the second term represents the spatial boundary condition taken for valence quarks; $\sigma_{val} = +1$ or $-1$ for the periodic ($P$) and the anti-periodic ($AP$) cases.

---
*presented by S. Aoki

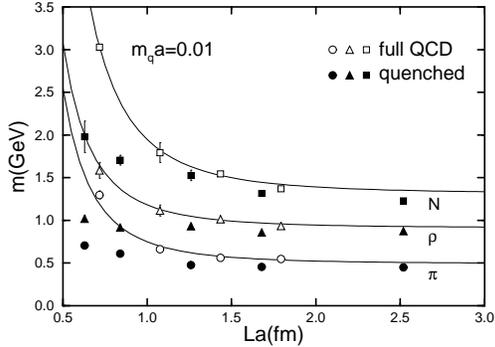

Figure 1. Comparison of quenched[1,4] and full QCD[3,6] hadron mass data. Solid lines are fits of form $m = m_\infty + c/L^3$ to full QCD data.

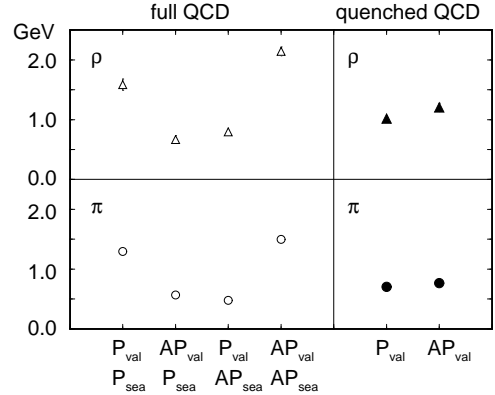

Figure 2. $\pi$ and $\rho$ masses for various quark boundary conditions at $m_q a = 0.01$.

It is obvious from (1) that hadron masses would be significantly shifted if $\langle P(C) \rangle$ takes a large value. This in fact is what happens in full QCD for small lattice sizes where breaking of center $Z(3)$ symmetry induced by dynamical sea quarks aligns Polyakov-type loops in some particular direction. In contrast, a sufficiently large ensemble of quenched configurations is invariant under $Z(3)$, and hence $\langle P(C) \rangle = 0$ even for small lattices unless the winding number of the loop is an integer multiple of 3. Thus quenched hadron masses should suffer less from finite-size effects, as is indeed observed in Fig. 1.

This line of argument can be pushed a step further and predict the sign of the mass shift for full QCD. If one imposes the anti-periodic boundary condition for sea quarks the Polyakov-type loops take positive values on the average, while for the periodic boundary condition they take negative values. Combined with the sign factor $\sigma_{val}$ due to valence quarks this means that the Polyakov-type terms cancel the Wilson-type contribution in (1) for the boundary conditions ($P_{sea}, P_{val}$) and ($AP_{sea}, AP_{val}$). Meson propagators will fall off faster for these cases, which would lead to an increase of meson masses. Similarly a decrease is expected for the opposite case of the ($P_{sea}, AP_{val}$) and ($AP_{sea}, P_{val}$) boundary conditions since the two terms add up.

In Fig. 2 we plot meson mass data taken on an $8^3 \times 16$ lattice with the four boundary conditions, which confirm these expectations. In the figure we also plotted quenched results for the $P_{val}$ and $AP_{val}$ boundary conditions on a $6^3 \times 40$ lattice. A small difference seen between the two cases for quenched QCD, negligible compared to those for full QCD, provides confirmation that Polyakov-type terms are almost cancelled out in quenched QCD as argued above.

An interesting way to further check the cancellation of Polyakov-type loops for quenched QCD is to impose a bias on their value. For example, placing a cut $Re \sum_{i=x,y,z} P_i < 0$ with $P_i$ the Polyakov loop in the $i$-th direction, one selects configurations which mimic those of full QCD generated with the $P$ boundary condition for sea quarks. For the $P$ boundary condition for valence quarks we then expect larger values of hadron masses. The results plotted in Fig. 3 show that the size-dependent shift of masses indeed become almost as large as those for full QCD.

## 4. Finite-size effect for large volume

Our results and analyses demonstrate that finite-size effects are smaller for quenched QCD. This, however, does not mean that a lattice size of order 2 to 3fm often used in quenched hadron mass measurements is large enough to ignore finite-size errors. This is illustrated in Fig. 4 where we plot our Kogut-Susskind results at $\beta = 6.0$ for the ratio $m_N/m_\rho$ extrapolated to the chiral limit as a function of $1/La$. The GF11 results for the Wilson action obtained with a Gaussian smeared sink of a size 0, 1 and 2 at $\beta = 5.7$[8] are also plotted. We observe an almost linear decrease of our data, amounting to 11% between $La \approx 1.7$fm to the largest lattice size $La \approx 2.5$fm. An 8% decrease of this ratio at $m_q a = 0.01$ is enhanced by the $m_q a \to 0$ extrapolation; this indicates an increasingly important finite-size effects towards the chiral limit. The decrease of the GF11 result extending to a larger lattice size $La \approx 3.3$fm is somewhat less but still significant. We should add that much less size dependence is found in the GF11 data for the sink size 4[8] and also in the previous APE data for Wilson action at the same $\beta$ and lattice sizes[9].

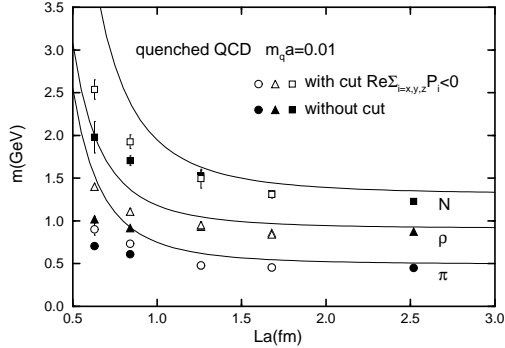

Figure 3. Comparison of quenched hadron masses with and without the cut $Re\sum_{i=x,y,z} P_i < 0$. Solid lines are fits to full QCD data from Fig.1.

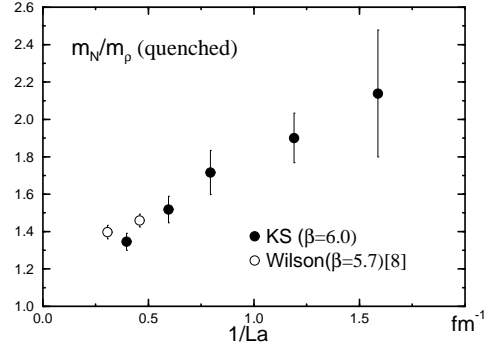

Figure 4. $m_N/m_\rho$ in the chiral limit as a function of $1/La$(fm).

## 5. Summary

We have shown that finite-size effects for small lattice sizes are much severer for full QCD. The difference originates from breaking of the center $Z(3)$ symmetry due to dynamical sea quarks in full QCD, which enhances the amplitude for propagation of valence quarks around the lattice. In quenched QCD the $Z(3)$ symmetry of the pure gauge action eliminates such amplitudes, and hence leads to a smaller finite-size effect.

Toward large lattice sizes Lüscher's analysis for point particles in a finite box leads us to expect that finite-size shift of hadron masses becomes exponentially small. Onset of such a behavior is not seen in current quenched data up to $La \approx 2.5 - 3.3$fm, however. Simulations on several different lattice sizes are necessary to extract the reliable value of $m_N/m_\rho$ at infinite volume.

## REFERENCES

1. S. Aoki et al., preprint UTHEP-265 (1993).
2. D. Toussaint, Nucl. Phys. **B**(Proc. Suppl.) **26** (1992) 3.
3. M. Fukugita et al., Phys. Rev. D**47** (1993) 4739.
4. N. Ishizuka et al., KEK-TH-351 (1993).
5. M. Fukugita et al., Phys. Lett. **B294** (1992) 380.
6. F. R. Brown et al., Phys. Rev. Lett. **67** (1991) 1062.
7. P. Hasenfratz and I. Montvay, Phys. Rev. Lett. **50** (1983) 209; G. Martinelli et al., Phys. Lett. **122B** (1983) 94; R. Gupta and A. Patel, Phys. Lett. **124B** (1983) 94.
8. F. Butler et al., Phys. Rev. Lett. **70** (1993) 2849 and preprint to appear.
9. P. Bacilieri et al., Nucl. Phys. **B317** (1989) 509; S. Cabasino et al., Nucl. Phys. **B**(Proc. Suppl.) **17** (1990) 431.